\documentclass[oneside,onecolumn,10pt]{IEEEtran}%

\usepackage{amsmath}
\usepackage{amsthm}
\usepackage{amsfonts}
\usepackage{amssymb}
\usepackage{graphicx}
\usepackage{color}
\usepackage{booktabs}
\usepackage{algorithmic}
\usepackage{algorithm}

\theoremstyle{plain}
\newtheorem{mytheorem}{Theorem}
\newtheorem{mycorollary}{Corollary}
\newtheorem{mylemma}{Lemma}

\theoremstyle{definition}
\newtheorem{obs}{Observation}

\renewcommand{\vec}[1]{\boldsymbol{#1}}

\newcommand{\set}[1]{\mathcal{#1}}

\newcommand{\Exx}[2]{\mathbb{E}_{{#1}}\left[ {#2} \right]}
\newcommand{\Reals}{\mathbb{R}}
\newcommand{\multi}[2]{\genfrac{}{}{0pt}{}{#1}{#2}}

\hypersetup {
  pdftitle          = {Nearly Doubling the Throughput of Multiuser MIMO Systems Using
Codebook Tailored Limited Feedback Protocol},
  pdfauthor         = {G. Wunder, J. Schreck, P. Jung },
  pdfsubject        = {},
  pdfkeywords       = {},
  bookmarksnumbered = true,   }

\newlength{\drop}

\newcommand*{\titleUL}{\begingroup
\drop=0.1\textheight
\vspace*{0.5\drop}
\begin{center}
{\LARGE\textsc{$\quad$}}\\[0.5\drop]
{{\LARGE IEEE Transactions on Wireless Communications} \\\vspace*{0.4cm} -- accepted for publication --  }\\[\drop]
\rule{\textwidth}{1pt}\par
\vspace{0.5\baselineskip}
{\huge\bfseries Nearly Doubling the Throughput of Multiuser \\ \vspace*{0.4cm}
  MIMO  Systems Using
Codebook Tailored \\ \vspace*{0.4cm} Limited Feedback Protocol
}\\[0.5\baselineskip]
\rule{\textwidth}{1pt}\par
\vfill
{\Large\textsc{Gerhard Wunder$^1$, Jan Schreck$^2$ and Peter Jung$^2$}}
\vfill
$^1$Fraunhofer Heinrich Hertz Institute, Einsteinufer 37,
  D-10587 Berlin, Germany\\
$^2$ Technische Universit\"at Berlin
Lehrstuhl f\"ur Informationstheorie und Theoretische Informationstechnik,
Einsteinufer 25, D-10587 Berlin, Germany  
\vfill
\today
\vfill
{\itshape \copyright 2012 IEEE. Personal use of this material is permitted. Permission from IEEE must be obtained for all other 
uses, in any current or future media, including reprinting/republishing this material for advertising or promotional purposes, creating new collective works, for resale or redistribution to servers or lists, or reuse of any 
copyrighted component of this work in other works.}
\end{center}
\endgroup}

\begin{document}
\titleUL
\thispagestyle{empty} 
\newpage
\begin{abstract}
We present and analyze a new robust feedback and transmit strategy for multiuser MIMO downlink
communication systems, termed Rate Approximation (RA). 
RA combines the flexibility and robustness needed for reliable communications with
the user terminal under a limited feedback constraint. It responds to two important observations.
First, it is not so significant to approximate the channel but rather the
rate, such that the optimal scheduling decision can be mimicked at the base station.
Second, a fixed transmit codebook at the transmitter is often better when
therefore the channel state information is more accurate. 
In the RA scheme the transmit and feedback codebook are 
separated and user rates are delivered to the base station subject to a controlled uniform error.
The scheme is analyzed and proved to have better performance below a certain interference plus noise margin
and better behavior than the classical Jindal formula.
LTE system simulations sustain the analytic results showing performance gains
of up to $50\%$ or $70\% $ compared to zeroforcing when using multiple
antennas at the base station and multiple antennas or a single antenna at the
terminals, respectively. 
A new feedback protocol is developed which inherently considers the transmit codebook
and which is able to deal with the complexity issue at the terminal. 
\end{abstract}
\section{Introduction}

Multiuser multiple input multiple output (MU--MIMO) communication systems have
been in the focus of intensive research over many years. The optimal
transmission technique for these systems is dirty paper coding (DPC), which,
under perfect channel state information at the transmitter (CSIT), achieves
superior performance gains over linear schemes. However, in practical systems
CSIT is obtained via a rate--constrained feedback channel, which is known to
be a sensitive part of the overall system and must be carefully designed.

In this paper we revisit the limited feedback problem in MU--MIMO systems. We
consider linear beamforming and assume that the transmit beamforming vectors
are defined by a fixed transmit codebook known to the base stations and all
users. In contrast to previous work, we use a different codebook for the
feedback and apply a new feedback strategy which we call \textit{Rate
Approximation} (RA). Loosely speaking, using the proposed RA feedback strategy,
the terminal selects a channel quantization vector from the feedback codebook
considering any possible scheduling decision that can be taken by the base
station. As we show, this will enable the base station \textit{to approximate
the user rates} (rather than the \textit{user channels}) subject to a small
uniform a priori error. Then, given the feedback message, the base station is
permitted to assert any beamforming vector from the transmit codebook for some
network oriented optimization purpose (not just the beamforming vector dictated by the user).


\subsection{Related Work}

An extensive survey on limited feedback in wireless multiple antennas systems
can be found in \cite{Love2008}. The standard reference for point--to--point
multiple input single output (MISO) systems is \cite{KiranMukkavilli2003}
where groundbreaking analytical expressions for the problem are derived.
Reference \cite{Au-Yeung2007a} evaluates the performance of point--to--point
MISO systems using random vector quantization (RVQ). In \cite{Santipach2009}
is shown that RVQ is asymptotically optimal for point--to--point MIMO systems.

For MU--MIMO systems, which are in the focus of this paper, reference
\cite{Jindal2006} provides the standard performance analysis for the
throughput degradation assuming RVQ.  In \cite{Kountouris07} different
feedback schemes are proposed that enable the base station to estimate the
signal--to--noise--plus--interference ration (SINR) of each user. However,
both papers specifically assume zeroforcing (ZF) beamforming and no individual
user rate analysis is provided. Reference \cite{Trivellato2008a} also
considers ZF and jointly designs the receive filters and the channel
quantization to maximize the expected SINR of each user. In \cite{Ding2007}
different kinds of partial CSIT are assumed and the performance of DPC and ZF
is compared.

Another popular transmission technique is unitary beamforming (UB). UB with a
sum feedback rate constraint is considered in \cite{Huang2007}. In
\cite{Samsung2006} a UB scheme named per user unitary rate control (PU2RC) has
been proposed for LTE. In \cite{Zhang2009} an improved user selection scheme
for PU2RC is proposed. Again, no error analysis for individual user rates has
been presented. 

In contrast to previous work, this paper considers an arbitrary transmission
scheme for a fixed transmit codebook and analyzes the individual rate error. 

\subsection{Organization and Main Results}

In Section \ref{sec:sysmodel} we introduce the system model and in Section
\ref{sec:RateApprox} the RA scheme is introduced. In Section \ref{sec:ana}:

\begin{enumerate}
\item We analyze the a priori rate error at the base station (before any
  scheduling decision) for each individual terminal evoked by our RA feedback
  strategy. We prove that it has better scaling properties compared to the
  classical result in \cite{Jindal2006} and that this benefit improves with an
  increasing number of transmit antennas.

\item We outline an advanced vector quantization problem related to the RA
  scheme by replacing the common chordal distance with a new distance function
  which inherently uses the structure of the transmit codebook.
\end{enumerate}

In Section \ref{sec:sim} we underline our results with LTE system simulations
showing the benefit obtained by the proposed RA scheme and develop a
suboptimal feedback protocol dealing with the complexity issue. This feedback
protocol is proposed to replace the common approach for LTE. Finally, in Sec.
\ref{sec:con} the conclusion is drawn with emphasis on the impact on future
standards.

\textbf{Notation:} Bold letters denote vectors and bold capital letters
matrices. The inner product between vectors $\vec{a}$ and $\vec{b}$ is defined
as $\langle\vec{a},\vec{b}\rangle=\vec{a}^{H}\vec{b}$, where $\vec{a}^{H}$ is
the conjugate transpose of the vector $\vec{a}$. The (euclidean) $\ell_{2}%
$-norm is $\lVert\vec{a}\rVert_{2}:=\langle\vec{a},\vec{a}\rangle^{1/2}$.
$\mathbb{S}^{n-1}$ is the unit sphere in $\mathbb{C}^{n}$. The $\ell_{1}%
$--norm of a vector $\vec{a}$ with components $a_{j}$ is defined as
$\lVert\vec{a}\rVert_{1}:=\sum_{j}|a_{j}|$ and $\lVert\vec{a}\rVert_{\infty
}:=\max_{j}|a_{j}|$ denotes its $\ell_{\infty}$--norm.

\section{System Setup}

\label{sec:sysmodel}

We consider the MU--MIMO downlink channel of a cellular system where a base
station, equipped with $n_{t}$ transmit antennas, serves multiple users,
equipped with $n_{r}$ receive antennas, on the same time and frequency
resource with a single data stream. The users are collected in the set
$\mathcal{U}$. Let $\vec x\in\mathbb{C}^{n_{t}}$ be the signal transmitted by
the base station in a single transmission interval (time index omitted). User
$m$ receives the transmitted signal through the channel $\vec H_{m}%
\in\mathbb{C}^{n_{r}\times n_{t}}$ and applies a fixed receive filter $\vec
u_{m}\in\mathbb{C}^{n_{r}}$ to recover its intended signal,
\[
y_{m}=\langle\vec u_{m}, \vec H_{m}\vec x\rangle+ n_{m} =:\langle\vec{\hat{h}%
}_{m},\vec{x}\rangle+n_{m}%
\]
where $n_{m}\sim\mathcal{C}\mathcal{N}(0,\sigma^{2})$ is additive white
Gaussian noise (zero--mean with variance $\sigma^{2}$) and $\vec{\hat{h}}$ is
the effective channel vector from the base station to user $m$. In the sequel
we assume each user $m$ has perfect knowledge of its own channel $\vec{{H}%
}_{m}$ and that the channels are constant over one transmission interval; no
fading model is imposed. Moreover, we assume no delay in the CSIT report,
scheduling or transmission.

In MU--MIMO systems adaptive adjustment of the number of active users is
crucial to achieve high spectral efficiency, see \cite{Zhang2011a} and
references therein. In each transmission interval the base station selects a
subset $\mathcal{S}\subseteq\mathcal{U}$ of users for transmission on the same
spectral resource and assigns each user $m\in\mathcal{S}$ a beamforming vector
out of a finite transmit codebook $\mathcal{C}\subset\mathbb{S}^{n_{t}-1}$,
known to the base station and all users. We will denote with $[\mathcal{C}%
]:=[1\dots|\mathcal{C}|]$ the set of codeword indices. The assignment of users
to beamforming vectors is defined by a mapping
\[
\pi:\mathcal{S}\rightarrow\lbrack\mathcal{C}],
\]
that maps each element $m\in\mathcal{S}$ to a codebook element $\vec{w}%
_{\pi(m)}\in\mathcal{C}$. We assume that $|\mathcal{S}|\leq n_{s}\leq n_{t}$,
where $n_{s}$ is the maximum number of users that can be scheduled on a
spectral resource. Note that we do not state the domain of $\pi$ explicitly, if it is
clear from the context. In the sequel we may assume that the codebook has the
property\footnote{This condition means that the codebook constitutes a
\emph{tight frame} for $\mathbb{C}^{n_{t}}$ with frame constant $A$} that for
every $\vec{f}\in\mathbb{C}^{n_{t}}$
\begin{equation}
   \sum_{\vec{w}\in\mathcal{C}}|\langle\vec{w},\vec{f}\rangle|^{2}=A\lVert\vec
   {f}\rVert_{2}^{2},
   \label{eq:tightframe}
\end{equation}
with a fixed constant $A\geq 1$. If $A=1$, $\mathcal{C}$ constitutes an
orthonormal base (ONB) and we call $\mathcal{C}$ an unitary codebook 
(used for UB).

Define the complex information symbols intended for user $m$ as $d_{m}
\in\mathbb{C}$, the transmitted signal is given by the superposition
\[
\vec{x}=\sqrt{\frac{P}{|\mathcal{S}|}}\cdot\sum_{m\in\mathcal{S}}\vec{w}%
_{\pi(m)}d_{m},
\]
where we assumed equal power allocation with the power budget $P$.
The achieved sum rate  for some user set 
$\mathcal{S}$ and mapping $\pi$ is
\[
  R\left(  \pi,\mathcal{S},H\right)  =\sum_{m\in\mathcal{S}}r_{m}(\pi
,\mathcal{S},\vec{\hat{h}}_{m})
\]
where $H=\{\vec{\hat{h}}_{m}\}_{m\in\mathcal{U}}$ is the list of effective
channels. The per user contributions to the sum rate are given by the Shannon
rates
\begin{equation*}
r_{m}(\pi,\mathcal{S},\vec{\hat{h}}_{m}):= \\ \log\biggl(1+\frac{|\langle\vec
{\hat{h}}_{m},\vec{w}_{\pi(m)}\rangle|^{2}}{\sigma^{2}|\mathcal{S}
|/P+\sum_{l\in\mathcal{S}\setminus\{m\}}|\langle\vec{\hat{h}}_{m},\vec{w}
_{\pi(l)}\rangle|^{2}}\biggr)
\end{equation*}
Throughout the paper we assume maximum sum rate scheduling, for instance, with
perfect CSIT the optimal user set $\mathcal{S}_{H}$ and mapping
$\pi_{H}$ is given as
\begin{equation}
\left(  \mathcal{S}_{H},\pi_{H}\right)  =\underset{%
\genfrac{}{}{0pt}{}{\mathcal{S}\subseteq\mathcal{U}}{\pi:\mathcal{S}%
\rightarrow\lbrack\mathcal{C}]}%
}{\arg\max}\,R\left(  \pi,\mathcal{S},H\right)  .\label{eq:sched_perfectCSIT}%
\end{equation}
However, due to the rate--constrained feedback channel, the base station takes
its decisions based solely on \textit{partial CSIT}. Partial CSIT message of
each user $m\in\mathcal{U}$ contains channel direction information (CDI)
$\vec{\nu}_{m}\in\mathcal{V}$ which is an element of the feedback codebook
$\mathcal{V}\subset\mathbb{S}^{n_{t}-1}$ of size $|\mathcal{V}|=2^{B}$ and
channel quality information (CQI) given by a scalar $\vartheta_{m}%
\in\mathbb{R}$. The feedback codebook is a priori known to all users and the
base station. Moreover, CQI is perfectly transferred to the base station,
which is a typical assumption, see e.g. \cite{Jindal2006}.

If the beamforming vectors are restricted to a fixed codebook $\mathcal{C}$
the scheduling decision based on partial CSIT $V=\{\vartheta_{m}\cdot\vec{\nu
}_{m}\}_{m\in\mathcal{U}}$ of all users $m\in\mathcal{U}$ can be found by
solving
\begin{equation}
\left(  \mathcal{S}_{V},\pi_{V}\right)  =\underset{%
\genfrac{}{}{0pt}{}{\mathcal{S}\subseteq\mathcal{U}}{\pi:\mathcal{S}%
\rightarrow\lbrack\mathcal{C}]}%
}{\arg\max}\,R\left(  \pi,\mathcal{S},V\right)  ,\label{eq:partCSI_sched}%
\end{equation}
where $R\left(  \pi,\mathcal{S},V\right)  =\sum_{m\in\mathcal{S}}r_{m}%
(\pi,\mathcal{S},\vartheta_{m}\vec{\nu}_{m})$. Equation
\eqref{eq:partCSI_sched} is a combinatorial problem that can be solved either
by a brute force search over the user sets $\mathcal{S}\subseteq\mathcal{U}$,
with $|\mathcal{S}|\leq n_{s}$, and the mappings $\pi:\mathcal{S}%
\rightarrow\lbrack\mathcal{C}]$ or more efficiently in a greedy fashion
\cite{Dimic2005, Trivellato2008a}. Clearly, the decisions in
(\ref{eq:partCSI_sched}) should match with the optimal decision
(\ref{eq:sched_perfectCSIT}) as good as possible. This is the motivation for
the following RA scheme.

\section{Rate Approximation}

\label{sec:RateApprox}

\subsection{RA Key Inequality}

The key idea of the RA scheme is to minimize the worst case rate mismatch
between the individual user rates in \eqref{eq:sched_perfectCSIT} and \eqref{eq:partCSI_sched} a priori and
independent of the (unknown) scheduling decision. The feedback message is
selected to make this error as small as possible.

Consider any baseline transmit scheme with perfect CSIT and sum rate $R(H)$.
Define the average rate gap between the baseline transmit scheme and
beamforming based on a fixed codebook with perfect CSIT as
\[
\Delta R_{\text{CSIT}}:=\mathbb{E}_{{H}}\left[  {R\left(  H\right)
}-{R\left(  \pi_{H},\mathcal{S}_{H},H\right)  }\right]
\]
and the average rate gap between the real sum rates $R\left(  \pi
,\mathcal{S},H\right)  $ and the approximated (based on partial CSIT) sum
rates $R\left(  \pi,\mathcal{S},V\right)  $ for a given user set $\mathcal{S}$
and mapping $\pi$ as:
\[
\Delta R\left(  \pi,\mathcal{S}\right)  :=\mathbb{E}_{H}\left[  {R\left(
\pi,\mathcal{S},H\right)  }-{R\left(  \pi,\mathcal{S},V\right)  }\right]  .
\]
Now, the rate gap between the baseline transmit scheme with perfect CSIT and
beamforming based on a fixed codebook with partial CSIT can be bounded from
above by
\begin{align}
\Delta R &  =\mathbb{E}_{H}\left[  {R\left(  H\right)  }-{R\left(  \pi
_{V},\mathcal{S}_{V},H\right)  }\right]  \nonumber\\
&  =\Delta R_{\text{CSIT}}+\mathbb{E}_{H}\left[  {R\left(  \pi_{H}%
,\mathcal{S}_{H},H\right)  }-{R\left(  \pi_{V},\mathcal{S}_{V},H\right)
}\right]  \nonumber\\
&  =\Delta R_{\text{CSIT}}+\Delta R\left(  \pi_{H},\mathcal{S}_{H}\right)
+\mathbb{E}_{H}\left[  {R\left(  \pi_{H},\mathcal{S}_{H},V\right)  }-{R\left(
\pi_{V},\mathcal{S}_{V},H\right)  }\right]  \nonumber\\
&  \leq\Delta R_{\text{CSIT}}+\Delta R\left(  \pi_{H},\mathcal{S}_{H}\right)
-\Delta R(\pi_{V},\mathcal{S}_{V})\label{eq:boundMain}\\
&  \leq\Delta R_{\text{CSIT}} +2\cdot \mathbb{E}_{{H}}\Biggl[ \sum_{m\in
\mathcal{S}_{H}\cup\mathcal{S}_{V}}  \max_{\multi{\set S\in\mathcal{S}_{m}}{\pi:\mathcal{S}\rightarrow
[ \mathcal{C}]}}\left\vert r_{m}(\pi,\mathcal{S},\vec{\hat{h}}_{m})-r_{m}(\pi,\mathcal{S}%
,\vartheta_{m}\vec{\nu}_{m})\right\vert \Biggr]  ,\label{eq:RAmot}%
\end{align}
where \eqref{eq:boundMain} must hold since $\pi_{V}$ is the optimal mapping of
users to beamforming vectors under the channel state information $V$. In
\eqref{eq:RAmot} we defined the set of user selections with maximal
cardinality $n_{s}$
\[
\mathcal{S}_{m}:=\left\{  \mathcal{S}\subseteq\mathcal{U}\,|\,m\in
\mathcal{S}\,\text{and}\,|\mathcal{S}|\leq n_{s}\right\}  ,
\]
which include user $m$. Moreover, we exploited that the rate gap $\Delta
R\left(  \pi_{H}\right)  -\Delta R(\pi_{V})$ is bounded from above by the
worst case rate gap
\begin{equation}
  \Delta R_{\text{RA}}:= 
  2 \cdot \mathbb{E}_{H}\Biggl[  
  \sum_{m\in\mathcal{S}_{H}\cup\mathcal{S}_{V}} \\ 
  \max_{ \multi{\mathcal{S}\in\mathcal{S}_{m}}{\pi:\mathcal{S}\rightarrow\lbrack\mathcal{C}]}}
    \left\vert 
      r_{m}(\pi,\mathcal{S},\vec{\hat{h}}_{m}) - 
      r_{m}(\pi,\mathcal{S},\vartheta_{m}\vec{\nu}_{m})
    \right\vert \Biggr]  \label{eq:deltaRA}
\end{equation}
From \eqref{eq:RAmot} we observe the following strategy which is the
motivation for the RA scheme, described in the next subsection.

\begin{obs}
To control $\Delta R_{\text{RA}}$ each user needs to individually minimize the
individual rate gap $\left\vert r_{m}(\pi,\mathcal{S},\vec{\hat{h}}_{m}%
)-r_{m}(\pi,\mathcal{S},\vartheta_{m}\vec{\nu}_{m})\right\vert $ for any
\mbox{$\mathcal{S}\in\mathcal{S}_{m}$} and mapping $\pi:\mathcal{S}\rightarrow
\lbrack\mathcal{C}]$.
\end{obs}

\subsection{RA Feedback Scheme}

To determine its feedback message each user $m\in\mathcal{U}$ must find a
tuple $\left(  \vartheta_{m},\vec{\nu}_{m}\right)  \in\left(  \mathbb{R}%
,\mathcal{V}\right)  $ that minimizes the RA distance\footnote{A closer look
reveals that it is neither in all cases a distance on $\mathbb{C}^{n_{t}}$ nor
on the Grassmann manifold.}
\begin{equation}
d(\vec{x},\vec{y})=\max_{%
\genfrac{}{}{0pt}{}{S\in\mathcal{S}_{m}}{\pi:\mathcal{S}\rightarrow
\lbrack\mathcal{C}]}%
}|r_{m}(\pi,\mathcal{S},\vec{x})-r_{m}\left(  \pi,\mathcal{S},\vec{y}\right)
|.\label{eq:RAdist}%
\end{equation}
Hence, each user $m\in\mathcal{U}$ finds its feedback message by solving
\begin{equation}
\left(  \vartheta_{m},\vec{\nu}_{m}\right)  =\underset{%
\genfrac{}{}{0pt}{}{\vartheta\in\mathbb{R}}{\vec{\nu}\in\mathcal{V}}%
}{\arg\min}\,d(\vec{\hat{h}}_{m},\vartheta\vec{\nu}).\label{eq:RAmax}%
\end{equation}
The RA scheme can be easily extended to users with multiple receive antennas
$n_{r}>1$. In this case for each scheduling decision $\pi:\mathcal{S}%
\rightarrow\lbrack\mathcal{C}]$ the optimal receive filter can be considered
in the RA distance according to
\[
r_{m}(\pi,\mathcal{S},\lambda_{m}\vec{h}_{m})=\max_{\vec{u}\in\mathbb{C}%
^{n_{r}}}r_{m}\left(  \pi,\mathcal{S},\vec{H}_{m}^{H}\vec{u}\right)  .
\]

Although not apparent at this point let us indicate some relevant properties
of the RA scheme: first, in the RA distance $d(\cdot,\cdot)$ the transmit
codebook matters which seems good engineering practice as we use all the
available information. Second, the terminals provide an uniform error which
indicates how well the rates are approximated and leads to inherent
robustness. This becomes particularly beneficial in the LTE multi antenna case
where channel state information is averaged over the subcarriers (see
Simulations in Section \ref{sec:sim}). Third, the RA scheme is amendable to
codebook optimization based on the RA distance function \eqref{eq:RAdist}; in
\cite{Schreck2009} we presented a codebook optimization algorithm for the RA
scheme which is based on the Lloyd algorithm.

\begin{figure}[tb]
\centering\includegraphics[width=.7\linewidth]{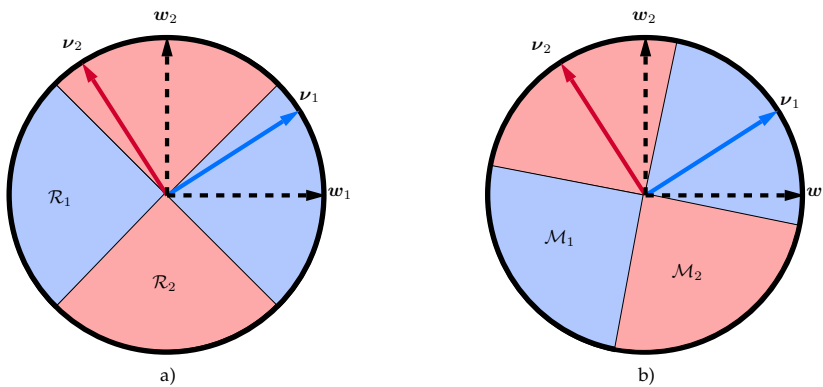}\newline\caption{ Toy
example in $\mathbb{R}^{2}$, with $n_{s}=2$, $n_{t}=2$ and $n_{r}=1$, hence,
$\vec{h}_{m}\in\mathbb{R}^{2}$. The transmit codebook (dashed black arrows) is
given by the columns of the identity matrix and the feedback codebook (solid
blues and red arrows) is given by a rotated version of the transmit codebook.
The CQI is equal to the receive SNR $\vartheta_{m}^{2}=\lambda_{m}^{2}$.
Comparing the feedback decisions taken under the RA distance a) and minimum
chordal distance b). The sets $\mathcal{R}_{i}$ and $\mathcal{M}_{i}$ show the
channel directions that result in feeding back $\nu^{[i]}$ under the RA
distance and the choral distance, respectively. }%
\label{fig:dis_reg2}%
\end{figure}
Finally, consider the example in Figure \ref{fig:dis_reg2} which establishes
that the RA distance indeed yields different feedback decisions compared to
the standard chordal distance \cite{Jindal2006}%
\begin{equation}
d_{C}(\vec{h}_{m},\vec{\nu})=\sqrt{1-|\langle\vec{h}_{m},\vec{\nu}\rangle
|^{2}}.\label{eq:MD}%
\end{equation}
The RA scheme's feedback decisions is obviously more oriented towards the
transmit codebook. In the following we analyze its performance. Moreover, as a
by-product a simpler distance is derived which is easier to calculate than the
computationally complex RA distance.

\section{Performance Analysis}

\label{sec:ana}

\subsection{Benchmarking Strategy}

Many papers prove that a particular transmission scheme achieves the optimal
multiuser multiplexing gain. That is, for sufficiently large $|\mathcal{U}|$
the sum rate scales like $n_{t}\log\log|\mathcal{U}|$. For instance this was
shown for random beamforming \cite{Sharif2005}, ZF \cite{Yoo07}, UB
\cite{Huang09} and RA \cite{Wunder2010a}. However, since rates and the number
of users are finite in a practical system, the significance of these
asymptotic results can at least be questioned. Putting it the other way
around: two methods achieving the optimal gain might behave completely
different in a practical system.

Our analysis is different and more inspired by the finite user results in
\cite{Jindal2006} and \cite{Francisco2007}. We assume: the number of users is equal to
the number of transmit antennas $|\mathcal{U}|=n_{t}$, all users are active
$\mathcal{S}=\mathcal{U}$ and the transmit codebook $\mathcal{C}$ constitutes
an ONB which corresponds to UB. This assumptions enable stringent comparison
to Jindal's result in \cite{Jindal2006} with ZF beamforming. Later, in Section
\ref{sec:US} we consider also user selection $\mathcal{S}\subseteq\mathcal{U}$
and general codebooks.

In the remainder of this section we will solely evaluate $\Delta R_{\text{RA}%
}$ in \eqref{eq:deltaRA}. The term $\Delta R_{\text{CSIT}}$ in
\eqref{eq:RAmot} was analyzed in \cite{Wunder10} and \cite{Francisco2007}, where it is
shown that for a certain SNR range (in the low SNR regime) $\Delta
R_{\text{CSIT}}$ can be even negative.

\subsection{Uniform RA Error with UB}

\label{sec:deltaRA}

For the ease of presentation define the normalized effective channel $\vec
{h}_{m}=\vec{\hat{h}}_{m}/\Vert\vec{\hat{h}}_{m}\Vert_{2}$ and the receive SNR
(normalized to the number $n_{t}$ of transmit antennas) of user $m\in
\mathcal{U}$ as%
\[
\lambda_{m}^{2}:=\frac{P\Vert\vec{\hat{h}}_{m}\Vert_{2}^{2}}{n_{t}\sigma^{2}}.
\]
Let us first provide a general expression for the maximum in
\eqref{eq:deltaRA} which gives us a hint how the RA scheme operates. Note that
when the RA scheme operates on a unitary transmit codebook we will denote this
scheme by RA--UB.

\begin{mylemma}
\label{lemma:RAvsID} If $\mathcal{U}=\mathcal{S}=\{1,2,\ldots,n_{t}\}$ and
$\mathcal{C}\subseteq\mathcal{V}$ then for some pair $(\vec{h}_{m},\vec{\nu})$
under the RA--UB scheme equation \eqref{lemma1} holds, 
\begin{equation}\label{lemma1}
d(\vec{\hat{h}}_{m},\vartheta\vec{\nu})\\
\leq{\max_{\vec{w}\neq\vec{w}^{\ast}}}\log\left(  1+\frac{\lambda_{m}%
^{2}\left(  \left\vert |\langle\vec{h}_{m},\vec{w}\rangle|^{2}-|\langle
\vec{\nu},\vec{w}\rangle|^{2}\right\vert +\frac{|\langle\vec{\nu},\vec
{w}\rangle|^{2}}{|\langle\vec{\nu},\vec{w}^{\ast}\rangle|^{2}}\left\vert
|\langle\vec{\nu},\vec{w}^{\ast}\rangle|^{2}-|\langle\vec{h}_{m},\vec{w}%
^{\ast}\rangle|^{2}\right\vert \right)  }{1+\lambda_{m}^{2}\left(
1-\max\left\{  |\langle\vec{h}_{m},\vec{w}\rangle|^{2},|\langle\vec{\nu}%
,\vec{w}\rangle|^{2}\right\}  \right)  }\right)
\end{equation}
where we defined $\vec{w}^{\ast}$ by $|\langle\vec{h}_{m},\vec{w}\rangle
|^{2}\leq|\langle\vec{h}_{m},\vec{w}^{\ast}\rangle|^{2}$ for all $\vec{w}%
\in\mathcal{C}$. The strategy minimizing the upper bound is to pick $\vec{\nu
}$ close to $\vec{h}_{m}$ (in the chordal distance) constrained by
$|\langle\vec{\nu},\vec{w}^{\ast}\rangle|^{2}\geq|\langle\vec{h}_{m},\vec
{w}^{\ast}\rangle|^{2}$.
\end{mylemma}

\begin{figure}[tb]
\centering\includegraphics[width=.5\linewidth]{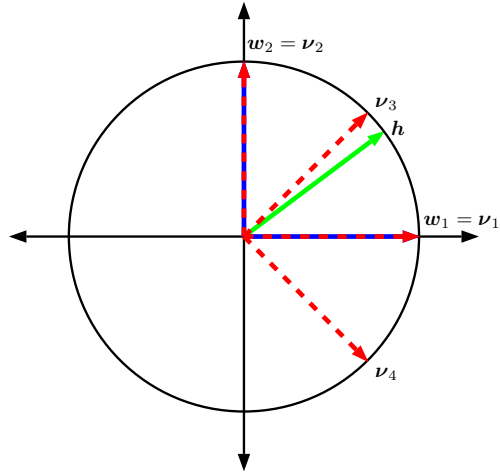}
\caption{
Example in $\mathbb{R}^{2}$, with $n_{t}=2$ and $n_{r}=1$. The suboptimal
feedback strategy considered in Lemma \ref{lemma:RAvsID} selects $\vec{\nu
}_{1}$ since $|\langle\vec{\nu}_{1},\vec{w}^{\ast}\rangle|\geq|\langle\vec
{h},\vec{w}^{\ast}\rangle|$ must hold and in this example $\vec{w}^{\ast}%
=\vec{w}_{1}$. By contrast, the minimum chordal distance \eqref{eq:MD} selects
$\vec{\nu}_{3}$ since $d_{C}(\vec{h},\vec{\nu})$ is minimized by $\vec{\nu
}=\vec{\nu}_{3}$.}
\label{fig:lemma1}
\end{figure}
The proof can
be found in Appendix \ref{proof:RAvsID}.
\begin{obs}
The suboptimal strategy in Lemma \ref{lemma:RAvsID} supports the intuition
that the error for the best beamformer should be set to zero by the applied
feedback strategy, see the example in Figure \ref{fig:lemma1}. 
\end{obs}

So far we are not able to effectively bound $\Delta R_{\text{RA}}$ which is
now settled based on Lemma \ref{lemma:RAvsID}. The following lemma shows that
$\Delta R_{\text{RA}}$ remains bounded when the SNR increases and that the
rate error depends solely on the function
\begin{equation*}
D_{m}(B):=\min_{\mathcal{V},|\mathcal{V}|=2^{B}}\mathbb{E}_{H}\Biggl[
\frac{1}{1-\tilde{\lambda}_{m}} \cdot \\  \min_{
\genfrac{}{}{0pt}{}{1>\tilde{\vartheta}_{m}>0}{\vec{\nu}\in\mathcal{V}}
}\max_{\vec{w}\in\mathcal{C}}\left\vert \tilde{\lambda}_{m}|\langle\vec{h}
_{m},\vec{w}\rangle|^{2}-\tilde{\vartheta}_{m}|\langle\vec{\nu},\vec{w}
\rangle|^{2}\right\vert \Biggr]  ,
\end{equation*}
where we defined $\tilde{\lambda}_{m}=\frac{\lambda_{m}^{2}}{1+\lambda_{m}%
^{2}}$ and $\tilde{\vartheta}_{m}=\frac{\vartheta_{m}^{2}}{1+\vartheta_{m}%
^{2}}$ in the proof of Lemma \ref{lemma:RAvsID} and $B$ is the number of
feedback bits.

\begin{mylemma}
\label{lem:deltaRAbounded} If $\mathcal{U}=\mathcal{S}=\{1,2,\ldots,n_{t}\}$
and $\mathcal{C}\subseteq\mathcal{V}$ then under the RA--UB scheme
\[
\Delta R_{\text{RA}}\leq2\sum_{m=1}^{n_{t}}\log\left(  1+\min_{\epsilon
>0}\frac{\left(  1+\epsilon\right)  D_{m}(B)}{1+\frac{\epsilon}{n_{t}-1}%
D_{m}(B)}\right)
\]

\end{mylemma}

The proof can be found in Appendix \ref{proof:deltaRAbounded}. The following
lemma gives a fundamental bound on $D_{m}(B)$.

\begin{mylemma}
\label{lem:d(b)} If the transmit codebook
$\mathcal{C}$ is unitary, then
\[
D_{m}(B)\leq c(n_{t})\mathbb{E}_{{H}}\left[  {\lambda_{m}^{2}}\right]
2^{-\frac{B}{n_{t}-1}},
\]
with
\begin{equation}
c(n_{t})=\left(  \Theta(\mathcal{B}_{2}^{n_{t}-1})\binom{2n_{t}-2}{n_{t}%
-1}\frac{\Gamma(1+\frac{n_{t}-1}{2})\sqrt{n_{t}}}{(n_{t}-1)!\pi^{\frac
{n_{t}-1}{2}}}\right)  ^{\frac{1}{n_{t}-1}}\label{eq:c(nt)}%
\end{equation}
and $B\geq\frac{(n_{t}-1)}{2}\log[(n_{t}-1)\sqrt{n_{t}-1}]$. For $n_{t}-1$
small tight bounds are known for the covering density $\Theta(\mathcal{B}%
_{2}^{n_{t}-1})$, e.g. $\Theta(\mathcal{B}_{2}^{2})\leq1.2091$ (Kershner,
1939), $\Theta(\mathcal{B}_{2}^{3})\leq1.4635$ (Bambah, 1954), $\Theta
(\mathcal{B}_{2}^{4})\leq1.7655$ (Delone \& Ryshkov, 1963). For $n_{t}-1\geq3$
the Rogers bound \cite{Boroczky2004} $\Theta(\mathcal{B}_{2}^{n_{t}%
-1})<4(n_{t}-1)\log(n_{t}-1)$ can be used.
\end{mylemma}

The complete proof can be found in the Appendix \ref{proof:d(b)}. Note that
$c(n_{t})$ is close to unity and falls below unity not before $n_{t} \geq 14$,
as required for improved scaling compared to Jindal's result. As the following
illustration for the case $n_{t} = 3$ shows, this is simply an artefact of the
proof technique.

\begin{figure}[tb]
\centering
\includegraphics[width =0.4\linewidth]{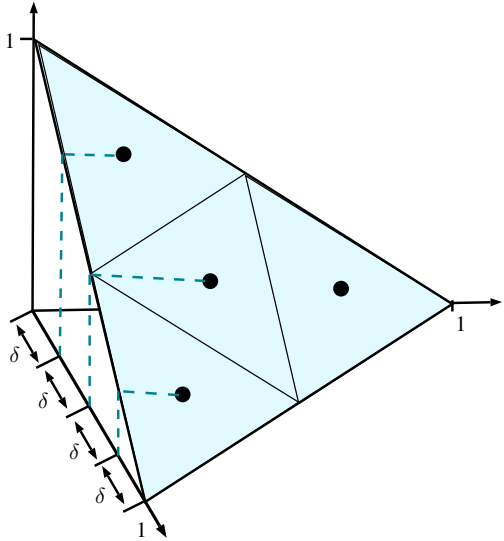}
\caption{The standard $2$--simplex in $3$ dimensions. The projection of the
quantization points $\mathcal{Q}$ on the coordinate axes implies a worst case
quantization error $\delta$. }%
\label{fig:voro}%
\end{figure}

Without loss of generality, we assume the unitary transmit codebook is given
by the standard ONB. We drop the user index $m$ and define from its channel
direction $\vec{h}$ the real positive vectors $\vec\psi=\left( \psi_{1},
\ldots, \psi_{3}\right) $ with $\psi_{n}:=|\langle\vec{h},\vec{w}_{\pi
(n)}\rangle|^{2}$ and $\vec\phi^{\vec\nu} =\left( \phi^{\vec\nu}_{1}%
,\ldots,\phi^{\vec\nu}_{3}\right) $ with $\phi^{\vec\nu}_{n}:=|\langle\vec
{\nu},\vec{w}_{\pi(n)}\rangle|^{2}$ for each $\vec{\nu}\in\mathcal{V}$. Per
definition, all these vectors have unit $\ell_{1}$--norm, $\|\vec\psi
\|_{1}=\|\vec\phi^{\vec\nu}\|_{1}=1$ and, hence, define points on the standard
$2$--simplex. Further, $\max_{\pi}\left\vert \psi_{n}-\phi^{\vec\nu}%
_{n}\right\vert =\lVert\vec\psi-\vec\phi^{\vec\nu}\rVert_{\infty}$ defines a
distance between two points on the standard $2$--simplex. Hence, for a given
feedback codebook $\mathcal{V}$ we can define the Voronoi region around the
point $\vec\phi^{\vec\nu}$ for a particular $\vec\nu\in\mathcal{V}$ as
$V(\vec\phi^{\vec\nu}) = \{\vec x \in\mathbb{R}_{+}^{3}: \|\vec x - \vec
\phi^{\vec\nu} \|_{\infty}< \| \vec x - \vec\phi^{\vec\xi} \|_{\infty}, \,
\forall\vec{\xi} \in\mathcal{V},\, \vec\xi\neq\vec\nu\}. $ If $B\in
\{1,2,4,\ldots\}$ and $n_{t}=3$, the feedback codebook can be chosen such that
the Voronoi regions are $2$--simplices with edge length $\tilde\delta\leq
\sqrt{2}$. Now, using the symmetry of the covering and projecting the
quantization points back on the coordinate axes (see Figure \ref{fig:voro}) we
get $\max_{\vec x \in V(\vec\phi^{\vec{\nu}})}\|\vec x - \vec\phi^{\vec{\nu}}
\|_{\infty}=\tilde\delta/\sqrt{8} = \delta$.

Now we can compute the volumes of the $2$--simplices (the standard simplex and
the scaled simplex) and proceed as in the proof of Lemma \ref{lem:d(b)} to
obtain the result
\[
\delta=\max_{\vec{x}\in V(\vec{\phi}^{\vec{\nu}})}\Vert\vec{x}-\vec{\phi
}^{\vec{\nu}}\Vert_{\infty}=2^{-\frac{B}{n_{t}-1}-1}.
\]
Hence, if $n_{t}=3$ and Rayleigh fading is assumed (i.e. $\vec{\hat{h}}%
_{m,i}\sim\mathcal{C}\mathcal{N}(0,1)\ $for $i=1,\dots,n_{t}$ and
$m\in\mathcal{U}$), the rate loss due to the rate--constrained feedback
channel scales like
\begin{equation*}
\Delta R_{\text{RA}}\leq \\ 2\sum_{m=1}^{n_{t}}\log\left(  1+\min_{\epsilon
>0}\frac{\left(  1+\epsilon\right)  \mathbb{E}_{{H}}\left[  {\frac
{\tilde{\lambda}_{m}}{1-\tilde{\lambda}_{m}}}\right]  2^{-\frac{B}{n_{t}-1}%
-1}}{1+\mathbb{E}_{{H}}\left[  {\frac{\tilde{\lambda}_{m}}{1-\tilde{\lambda
}_{m}}}\right]  \frac{\epsilon}{n_{t}-1}2^{-\frac{B}{n_{t}-1}-1}}\right)
\\ \leq2\sum_{m=1}^{n_{t}}\log\left(  1+\frac{P}{\sigma^{2}}2^{-\frac{B}{n_{t}%
-1}-1}\right)  .
\end{equation*}
Therefore, we have an improvement of $n_{t}-1$ bits in the exponential term
compared to Jindal's result for ZF with feedback based on minimizing the
chordal distance (see \cite{Jindal2006}), under the very same assumptions.

\subsection{Uniform RA Error with User Selection and General Codebooks}

\label{sec:US} In this subsection we no longer assume unitary transmit
codebooks and allow user selection at the base station.

\begin{mytheorem}
\label{theo:US} Assuming Rayleigh fading, arbitrary transmit codebooks and
user selection $\mathcal{S}\subseteq\mathcal{U} = \{1,2,\ldots,n_{t}\}$, then
for any $\set V$ under the RA scheme 
\begin{equation*}
\Delta R_{\text{RA}} \leq  4n_{s} \log\biggl(  1+\frac{Pn_{t}}{\sigma^{2}%
} \\ \mathbb{E}_{{H}}\left[  {\min_{\vec{\nu}\in\mathcal{V}}\max_{\vec{w}%
\in\mathcal{C}}\left\vert |\langle\vec{h},\vec{w}\rangle|^{2}-|\langle\vec
{\nu},\vec{w}\rangle|^{2}\right\vert }\right]  \biggr)  .
\end{equation*}

\end{mytheorem}
The proof can be found in Appendix \ref{proof:US}. The expected value
\[
\hat{D}_{m}(B):=\min_{\mathcal{V},|\mathcal{V}|=2^{B}}\mathbb{E}_{{H}}\left[  {\min_{\vec{\nu}\in\mathcal{V}}%
\max_{\vec{w}\in\mathcal{C}}\left\vert |\langle\vec{h},\vec{w}\rangle
|^{2}-|\langle\vec{\nu},\vec{w}\rangle|^{2}\right\vert }\right]
\]
has been shown to be analytically tractable, in the previous section, for
unitary transmit codebooks. For codebooks constituting a tight frame 
(see the condition \eqref{eq:tightframe}) we devise the following corollary.

\begin{mycorollary}
If the transmit codebook $\mathcal{C}$ is a tight frame, then
$
\hat{D}_{m}(B)\leq A\cdot c(n_{t})2^{-\frac{B}{n_{t}-1}},
$
where $c(n_{t})$ is defined in \eqref{eq:c(nt)} and $A$ is the frame constant
in \eqref{eq:tightframe}.
\end{mycorollary}
The proof is a simple extension of Lemma \ref{lem:d(b)} and omitted. 
The previous result is remarkable since all the $2^{n_{t}}$ possible user
rates are uniformly recovered at the base station with better scaling
properties than the classical result. The RA scheme is now applied in a
practical scenario.

\section{Practical Considerations and Simulations}

\label{sec:sim}

\subsection{Efficient and Robust Feedback Protocol}

\label{sec:effFB} Mobile user equipments usually have limited computing
capabilities, therefore, most systems require that the complexity at the user
side is as low as possible. Hence, solving the full rate approximation problem
(the min--max problem \eqref{eq:RAmax}) may not be feasible. Fortunately, our
analysis in Section \ref{sec:ana} yields the (suboptimal) distance function
\begin{equation}
d_{S}(\vec{h},\vec{\nu})=\max_{\vec{w}\in\mathcal{C}}\left\vert \left\vert
\langle\vec{h},\vec{w}\rangle\right\vert ^{2}-\left\vert \langle\vec{\nu}%
,\vec{w}\rangle\right\vert ^{2}\right\vert ,\label{eq:RA_low}%
\end{equation}
which can be used at the user side to uniformly bound the rate approximation
error $\Delta R_{\text{RA}}$ \eqref{eq:deltaRA}. Further, we define the CQI
reported by user $m$ as
\begin{equation}
\vartheta_{m}^{2}=\lambda_{m}^{2}|\langle\vec{h}_{m},\vec{\nu}_{m}\rangle
|^{2},\label{eq:CQI}%
\end{equation}
which can be interpreted as the effective channel gain of user $m$ over the
quantized channel $\vec{\nu}_{m}$. Equation \eqref{eq:CQI} captures two
important aspects. On the one hand, if the CDI is equal to the channel
direction, the user gets no penalty ($|\langle\vec{h}_{m},\vec{\nu}_{m}%
\rangle|^{2}=1$) on the other hand if the CDI is orthogonal to the channel
direction, the effective channel is zero ($|\langle\vec{h}_{m},\vec{\nu}%
_{m}\rangle|^{2}=0$). Hence, the CQI \eqref{eq:CQI} reflects the receive SNR
and the quantization error, which is also in accordance with the results in
\cite{Yoo07}. In Algorithm \ref{alg1} the efficient feedback protocol is summarized.
\begin{algorithm}[htbp]
\caption{Efficient and robust feedback protocol}
\label{alg1}
\begin{algorithmic}[1]
\FOR{All users $m\in \set U$}
\STATE Estimate channels based on common pilots.
\STATE Compute CDI according to \eqref{eq:RA_low}.
\STATE Compute CQI according to \eqref{eq:CQI}.
\STATE Feedback CDI and CQI to base station.
\ENDFOR
\end{algorithmic}
\end{algorithm}

\subsubsection*{Complexity of the proposed feedback protocol}

\begin{figure}[tb]
\centering
\includegraphics[width =.5\linewidth]{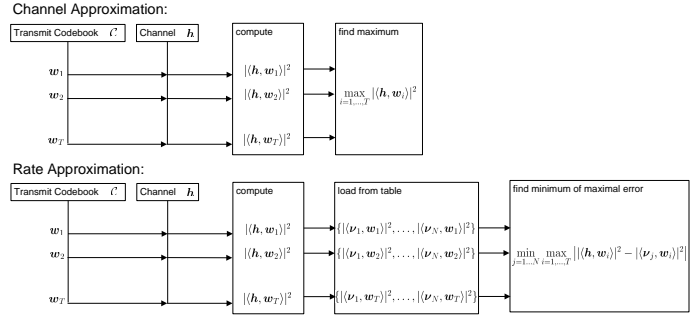}
\caption{Schematic comparison of the CDI computation at the user side; for the
efficient RA distance (bottom) and chordal distance (top). }%
\label{fig:flow}%
\end{figure}
Figure \ref{fig:flow} (bottom) shows a flow chart of the CDI computation with
the proposed feedback protocol using the distance function $d_{S}(\vec
h,\vec\nu)$ defined in \eqref{eq:RA_low}. We point out that the terms
$|\langle\vec\nu_{j} , \vec w_{i} \rangle|$, for $i=1,\ldots,|\mathcal{C}|$
and $j=1,\ldots,|\mathcal{V}|$, only need to be computed ones and can be
stored in the memory. Therefore, during the feedback phase user $m$ must only
compute $|\langle\vec h_{m} , \vec w_{i} \rangle|^{2}$ for all $i=1,\ldots
,|\mathcal{C}|$ and the difference $\bigl||\langle\vec h_{m} , \vec w_{i}
\rangle|^{2} - |\langle\vec\nu_{j} , \vec w_{i} \rangle|^{2}\bigr|$ for all
$i,j$. Figure \ref{fig:flow} (top) shows the steps that need to be performed
to compute the CDI based on the chordal distance \eqref{eq:MD}. To compute the
chordal distance each user must compute $|\langle\vec h_{m} , \vec\nu_{i}
\rangle|$ for all $i=1,\ldots,|\mathcal{V}|$.

\begin{table}[ptbh]
\caption{Number of scalar products that need to be evaluated to solve MD
\eqref{eq:MD} and the proposed RA distance \eqref{eq:RA_low}}%
\label{tab:complex}%
\centering
\begin{tabular}
[c]{|l|l|l|}\hline
B & MD \eqref{eq:MD} & proposed RA \eqref{eq:RA_low}\\\hline
1 & 2 & 16\\
2 & 4 & 32\\
3 & 8 & 64\\
4 & 16 & 128\\
8 & 256 & 2048\\\hline
\end{tabular}
\end{table}
If we assume a fixed transmit codebook $\mathcal{C}$ and a feedback codebook
$\mathcal{V}$ with $2^{B}$ elements, the complexity of computing the CDI based
on \eqref{eq:MD} or \eqref{eq:RA_low} is asymptotically equal, i.e., using
Landau notation $O(2^{B})$. However, this result is only valid for $2^{B}$
growing asymptotically large. For small values $B=1,2,3,4,8$ and
$|\mathcal{C}| = 8$ the number of scalar products that needs to be evaluated
are summarized in Table \ref{tab:complex}.

\subsection{Simulations}

In the simulations we consider a LTE like system architecture. That is,
multiple base stations transmit to multiple users using the spectrum. The
spectrum is divided in orthogonal subcarriers using orthogonal
frequency--division multiplexing OFDM. In the sequel we use a frequency reuse
factor of one, i.e., each base station uses the whole frequency band. Since,
we assume no cooperation between the base stations inter cell interference is
indispensable. In the sequel the channel from base station $b$ to user $m$ on
subcarrier $f$ is given by $\vec H_{m,b}(f)\in\mathbb{C}^{n_{r}\times n_{t}}$.

The transmit protocol can be summarized as follows. 
First, each base station transmits orthogonal common pilots.
Then, each user quantizes and feeds back its channel state information. 
Based on the quantized channel state information each base station solves the
scheduling problem \eqref{eq:partCSI_sched}.  
Finally, dedicated (i.e. precoded) pilots are transmitted by all base
stations.

\begin{table}[ptbh]
\caption{Simulation Parameters}%
\label{tab:simpar}%
\centering
\begin{tabular}
[c]{|p{5cm}|l|}\hline
Parameter & Value/Assumption\\\hline
Number of base stations & $3$\\
Frequency reuse & full\\
Number of users $|\mathcal{U}|$ & $30$ (uniformly distributed)\\
Number of transmit antennas $n_{t}$ & $4$ (uncorrelated)\\
Number of receive antennas $n_{r}$ & $1$ or $2$ (uncorrelated)\\
Receiver type & maximum ratio combining\\
Maximum number of scheduled users per scheduling block $n_{s}$ & 4\\
Equivalent SNR & $153$ dB\\
LTE carrier frequency / bandwidth & $2$ GHz / $10$ MHz\\
Number of PRB & 50\\
Scheduling block size & $1$ PRB $= 12$ subcarrier\\
LTE channel model & SCME (urban macro)\\
Inter cell interference modeling & explicit\\\hline
\end{tabular}
\end{table}
The simulation parameters are given in Table \ref{tab:simpar}, they can be
summarized as follows. $3$ base stations located in $3$ adjacent cells and
$30$ users uniformly distributed over the network area; given by a radius of
$250$ meter around the center of the base stations. The physical layer is
configured according to LTE \cite{LTE_Gen}. The base station are equipped with
$n_{t}=4$ transmit antennas and each user is equipped with $n_{r}=1$ or
$n_{r}=2$ receive antenna (specified in the caption). The transmit codebook
and feedback codebook is given by the LTE codebook defined in \cite{LTE_Gen}
which has $N=16$ elements and, hence, requires $B=4$ bit to feedback back the
CDI. The channels are modeled by the spatial channel model extended (SCME)
\cite{SCME2} using the urban macro scenario.

In total $600$ subcarriers per base station are available. The subcarriers are
clustered in groups of $F=12$ subcarriers; one subcarrier group is denoted as
physical resource block (PRB). One PRB is the smallest scheduling unit. The
subcarrier indexes of PRB $p$ are collected in the index set $\mathcal{F}_{p}%
$. We define the average channel gain of PRB $p$ as $\sigma^{2}_{p} =
1/|\mathcal{F}_{p}| \sum_{f\in\mathcal{F}_{s}} \|\vec H_{b,m}(f)\|_{F}^{2}$,
where $\|\vec A\|_{F}$ is the Frobenius norm of matrix $\vec A$, and assume
that each user is assigned to that base station with maximal total average
channel gain $1/F \sum_{f=1}^{F} \sigma^{2}_{f}$. Each user reports one
feedback message per PRB to that base station it is assigned to.

Each of the base stations runs an independent local scheduler. In every
transmission interval up to $n_{s} = 2$ users can be scheduled by each base
station on every PRB. Scheduling is performed in a greedy fashion according to
\cite{Trivellato07}. For simplicity we assume no delay in the CSIT report,
scheduling, transmission or performance evaluation.

The performance is evaluated based on the network spectral efficiency which we
define by
$
\sum_{b=1}^{B} \sum_{p= 1}^{F} \sum_{m\in\mathcal{S}_{b,p}} \sum
_{f\in\mathcal{F}_{p}}\log(1 + \text{SINR}_{m}(f)),
$
where $\text{SINR}_{m,b}(f)$ is the SINR of user $m$ on subcarrier $f$ and
$S_{b,p}$ are the users scheduled by base station $b$ on PRB $p$.

In the simulation we compare four different feedback strategies.

\begin{enumerate}
\item Perfect (average) CSIT: the base station knows the channel averaged
over  all subcarriers perfectly, $\vec{\bar H}_{m} = \frac{1}{F}\sum_{f=1}^{F}
\vec H_{m,f}, $ where $\vec H_{m,f}$ is the channel of user $m$ on subcarrier
$f$.

\item Minimum chordal distance: user $m$ determines its CDI feedback by
minimizing the chordal distance \eqref{eq:MD} to the channel $\vec{\bar
h}_{m}  = \langle\vec u, \vec{\bar H}_{m}\rangle$, where $\vec u$ is chosen
to  maximize $|\langle\vec u, \vec{\bar H}_{m} \rangle|$.

\item Rate Approximation as described in Section \ref{sec:RateApprox} with
the  rates
\[
r_{m}\left(  \pi,\mathcal{S},\lambda_{m}\vec{h}_{m}\right)  =  \frac{1}{F}%
\sum_{f=1}^{F}r_{m}\left(  \pi,\mathcal{S},\lambda_{m}\vec{h}_{m,f}\right)  .
\]

\item Efficient Rate Approximation as described in Section \ref{sec:effFB},
where $\vec h_{m}$ is given by the average channel $\vec{\bar h}_{m}$ as
defined  for minimum chordal distance above.
\end{enumerate}

\begin{figure}[tb]
\centering
\includegraphics[width =.5\linewidth]{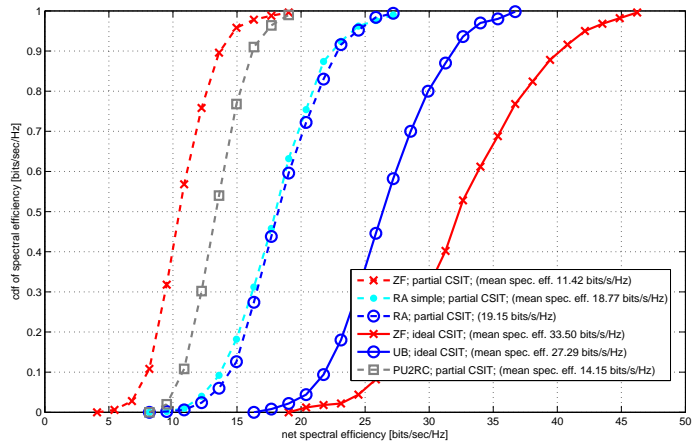}  \caption{System level
simulation: CDF of spectral efficiency for $n_{r} =1$; Comparing PU2RC, ZF, UB
and RA under ideal and partial CSIT}%
\label{fig:miso}%
\end{figure}
Figure \ref{fig:miso} depicts the CDF of the spectral efficiency for users
with $n_{r}=1$ receive antenna. The ZF scheme is implemented according to
\cite{Trivellato2008a}. The PU2RC scheme is based on the same transmit
codebook as RA and is implemented according to \cite{Samsung2006}. We observe
that with perfect CSIT ZF outperforms greedy scheduling with a fixed codebook.
With partial CSIT the RA scheme significantly outperforms ZF with a gain of
approximately $70 \%$. Remarkable is also the gain of about $35\%$ of RA over
PU2RC. Moreover, Figure \ref{fig:miso} shows that RA with the efficient
distance function \eqref{eq:RA_low} performs very close to the full RA scheme.

\begin{figure}[tb]
\centering
\includegraphics[width =.5\linewidth]{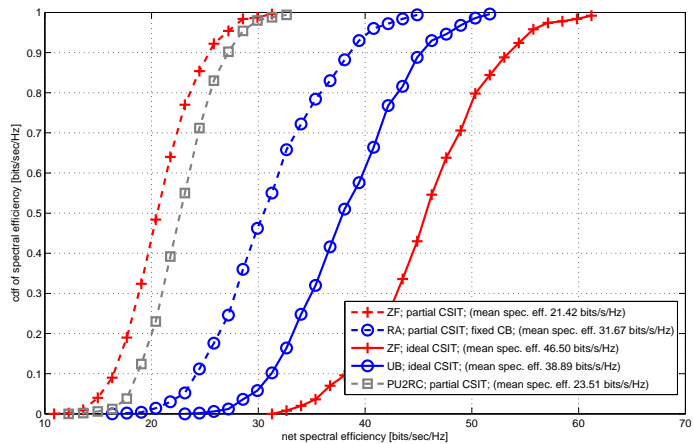}  \caption{System level
simulation: CDF of spectral efficiency for $n_{r} =2$; Comparing PU2RC, ZF, UB
and RA under ideal and partial CSIT}%
\label{fig:mimo}%
\end{figure}
Figure \ref{fig:mimo} depicts the CDF of the spectral efficiency for users
with $n_{r}=2$ receive antennas. We observe that with perfect CSIT ZF
outperforms greedy scheduling with a fixed codebook. With partial CSIT the RA
scheme significantly outperforms all other schemes and achieves a gain of
approximately $50 \%$ over ZF. Remarkable is also the $35 \%$ gain of RA over
PU2RC.

\begin{figure}[tb]
\centering
\includegraphics[width =.5\linewidth]{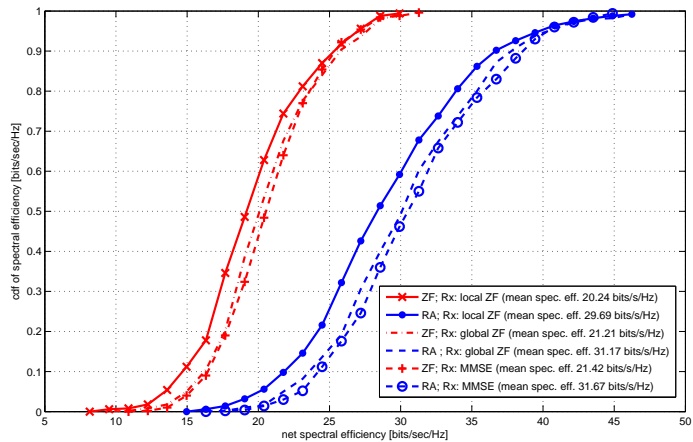}  \caption{System level
simulation: CDF of spectral efficiency for $n_{r} =2$; Comparing ZF and RA
with partial CSIT and different receive filters.}%
\label{fig:mimo_rx}%
\end{figure}
In Figure \ref{fig:mimo_rx} we compare the performance of RA and partial CSIT
ZF with different receive filters, i.e. the MMSE receive filter that maximizes the SINR of each user by considering the interference
from all other users, the global ZF receive filter that tries to minimize interference from
all base stations and the local ZF receive filter that considers only
interference from the own base station. We observe that both transmit schemes
achieve the highest network sum rate with the SINR optimal receive filter. The
performance degradation with the local and global ZF receive filter are
similar for both transmit schemes.

\section{Conclusion}

\label{sec:con} In this paper we invented and analyzed the rate approximation
scheme. It was shown that each user can individually minimize its rate error a
priori by selecting the feedback message in a robust fashion incorporating the
transmit codebook. The respective error expressions and feedback schemes are
derived and compared to the standard expressions. It is proved that a better
scaling is possible when the size of the transmit code book is small. A
remarkable result is that it is often much better to reduce flexibility at the
base station in favor of having more reliable CSIT.

\appendices

\section{Proof of Lemma \ref{lemma:RAvsID}}

\label{proof:RAvsID} \begin{IEEEproof}
Let us first drop the user index $m$, i.e. $\lambda=\lambda_m$ and $\vec{h}=\vec{h}_m$.
Further abbreviate $\vec{w}=\vec{w}_{\pi(m)}$, $\phi=|\langle \vec{\nu},\vec{w}\rangle|^2$ and
$\psi=|\langle \vec{h},\vec{w}\rangle|^2$ without explicitly writing the dependency of $\pi$.
Using $\set{U}=\set{S}$ with $|\set S|=n_{t}$ we get for any $\vartheta$ and
normalized vector $\vec{\nu}$
\begin{align}
r_{m}(\pi,\set S,\lambda\vec{h})-r_{m}(\pi,\set
  S,\vartheta\vec{\nu}) 
= & \log\left(  \frac{\left(  \lambda^{2}+1\right)  \left(  1+\vartheta^{2}\left(  1-\phi\right)  \right)  }{\left(
\vartheta^{2}+1\right)  \left(  1+\lambda^2\left(  1-\psi\right)  \right)  }\right) \\
  = & \log\left(  \frac{1-\tilde{\vartheta}\phi}{1-\tilde{\lambda}\psi}\right)
=\log\left(  1+\frac{\tilde{\lambda}\psi-\tilde{\vartheta}\phi}{1-\tilde{\lambda}\psi}\right)  .
\label{eq:zero}
\end{align}
Here, we have set
$\tilde{\vartheta}:=\frac{\vartheta^{2}}{\vartheta^{2}+1},\;\;\tilde{\lambda}:=\frac{\lambda^{2}}{\lambda^{2}+1}$.
Similarly, the negative term can be rewritten as
$-\log\left(  \frac{1-\tilde{\vartheta}\phi}{1-\tilde{\lambda}\psi}\right)
=\log\left(1+\frac{\tilde{\vartheta}\phi-\tilde{\lambda}\psi}{1-\tilde{\vartheta}\phi}\right)$.
Recall, that $\phi$ and $\psi$ depend on $\vec{w}$ which in turn depends again on the mapping $\pi$.
With the assumptions of this lemma we have that the set $\set{S}_m$ of possible scheduling subsets
in the definition of RA distance \eqref{eq:RAdist} is simple, i.e. $\set{S}_m=\{\set{U}\}$ and
we get from \eqref{eq:zero} the following upper bound
\begin{equation}
\begin{split}
d_m(\lambda\vec{h},\vartheta\vec{\nu})&=\max_{\pi} |r_{m}(\pi,\set S,\lambda\vec{h})-r_{m}(\pi,\set S,\vartheta\vec{\nu})|\\
&\leq
\max_{\pi}
\max\left[\log
\left(  1+\frac{\tilde{\lambda}\psi-\tilde{\vartheta}\phi}{1-\tilde{\lambda}\psi}\right)
,
\log
\left(1+\frac{\tilde{\vartheta}\phi-\tilde{\lambda}\psi}{1-\tilde{\vartheta}\phi}\right)
\right]\\
&\leq
\max_{\pi}\log\left(1+
\max\left[
\frac{\tilde{\lambda}\psi-\tilde{\vartheta}\phi}{1-  \tilde{\lambda}\psi}
,
\frac{\tilde{\vartheta}\phi-\tilde{\lambda}\psi}{1-\tilde{\vartheta}\phi}
\right]
\right)
\label{eq:first}
\end{split}
\end{equation}
which is still valid for any $\vec{\nu}\in\mathcal{V}$.
Now, consider the following (sub--optimal) three-step strategy for the feedback protocol:
(a) define $\vec{w}^*:=\arg\max_{\vec{w}\in\set C}|\langle\vec{h},\vec{w} \rangle|^{2}$ to be
the codeword nearest to true channel direction $\vec{h}$ in the chordal distance (see \eqref{eq:MD}) and
(b) select then a CDI $\vec{\nu}\in\mathcal{V}$ from the feedback codebook which
is closer to $\vec{w}^*$ as $\vec{h}$ is, i.e. which has the property
$\theta:=|\langle\vec{\nu},\vec{w}^{\ast}\rangle|^{2}\geq |\langle\vec{h},\vec{w}^{\ast}\rangle|^{2}=:\eta$.
Since $\mathcal{C}\subseteq\mathcal{V}$ such vector always
exists. (c) Determine the CQI $\vartheta$ by setting
$\frac{\lambda^{2}\eta}{1+\lambda^{2}(1-\eta)}=\frac{\vartheta^{2}\eta}{1+\vartheta^{2}(1-\theta)}$,
which yields after some calculations
$\tilde{\vartheta}=\frac{\lambda^{2}}{\lambda^{2}+1}\frac{\eta}{\theta}=\tilde{\lambda}\frac{\eta}{\theta}$.
The result of this strategy is that we get for the first term in the ''$\max$'' of \eqref{eq:first}:
\begin{align*}
\frac{\tilde{\lambda}\psi-\tilde{\vartheta}\phi}{1-\tilde{\lambda}\psi}
&\leq \tilde{\lambda}\cdot\frac{\psi-
\frac{\eta}{\theta}\phi}{1-\tilde{\lambda}\psi}
\leq\frac{\tilde{\lambda}}{1-\tilde{\lambda}\psi}\biggl(
|\psi-\phi|  +\frac{\phi}{\theta}|\theta-\eta|\biggr).
\end{align*}
for any $\vec{w}\neq\vec{w}^*$ (which ensures that $\psi=|\langle\vec{h},\vec{w}\rangle|^{2}<
|\langle\vec{h},\vec{w}^*\rangle|^{2}=\eta\leq1$).
Similar, for the second term in upper bound \eqref{eq:first} we obtain
\begin{align*}
\frac{\tilde{\vartheta}\phi-\tilde{\lambda}\psi}{1-\tilde{\vartheta}\phi}
&\leq\tilde{\lambda}\cdot
\frac{\frac{\eta}{\theta}\phi-\psi}{1-\tilde{\lambda}\frac{\eta}{\theta}\phi}
\overset{(\eta\leq\theta)}{\leq}
\tilde{\lambda}\cdot\frac{\frac{\eta}{\theta}\phi-\psi}{1-\tilde{\lambda}\phi}
\\ & \leq
\frac{\tilde{\lambda}}{1-\tilde{\lambda}\phi}\biggl(
|\psi-\phi|+\frac{\phi}{\theta}|\theta-\eta|\biggr)
\end{align*}
where we now need the additional property of step (b) that
$\eta=|\langle\vec{h},\vec{w}^{\ast}\rangle|^{2}\leq|\langle\vec{\nu},\vec{w}^{\ast}\rangle|^{2}=\theta$
which proves the claim.
\end{IEEEproof}

\section{Proof of Lemma \ref{lem:deltaRAbounded}}

\label{proof:deltaRAbounded} \begin{IEEEproof}
According to the rule \eqref{eq:RAmax}, RA aims on
minimizing the maximal rate error over the elements of $\mathcal{V}$.
Under the assumptions of this lemma (the list of scheduling subsets is  $\set{S}_m=\{\set{U}\}$)
the maximal rate error at user $m$ achieved for given CDI $\vec{\nu}_m$ and CQI $\vartheta_m$ is:
\begin{equation}
d_m(\lambda_{m}\vec{h}_{m},\vartheta\vec{\nu})
  = {\max_{\pi}|r_{m}(\pi,\set S,\lambda_{m}\vec{h}_{m})-r_{m}\left(
\pi,\set S,\vartheta_m\vec{\nu}_m\right)  |}
\end{equation}
and from \eqref{eq:deltaRA} we have in this case:
\begin{equation*}
\Delta R_{\text{RA}}:=2\sum_{m\in\set S}\mathbb{E}_{H}\left[d_m(\lambda_{m}\vec{h}_{m},\vartheta_m\vec{\nu}_m)\right]
\end{equation*}

Let us consider the contribution of user $m$ to this sum.
Using the notation of Appendix \ref{proof:RAvsID} we have from \eqref{eq:first}  and Jensen's inequality
that
\begin{equation}
\mathbb{E}_{H}\left[d_m(\lambda\vec{h},\vartheta\vec{\nu})\right] 
 \leq  \log\left(  1+\mathbb{E}_{H}
\left[  \min_{\multi{1>\tilde{\vartheta}_{m}>0}{\vec{\nu}
\in{\mathcal{V}}}}{\max_{\pi}}
\frac{|\tilde{\lambda}\psi-\tilde{\vartheta}\phi| }{1-\tilde{\lambda}\psi}\right]
\right)  +  
\log\left(  1+\mathbb{E}_{H}
\left[  \min_{\multi{1>\tilde{\vartheta}_{m}>0}{\vec{\nu}
\in{\mathcal{V}}}}{\max_{\pi}}
\frac{|\tilde{\lambda}\psi-\tilde{\vartheta}\phi|}{1-\tilde{\vartheta}\phi}\right]
\right).
\label{eq:ExpRVQ}
\end{equation}
Let us re-write the first term on the right side of \eqref{eq:ExpRVQ}. The
idea is to use Lemma  \ref{lemma:RAvsID} with $\set V = \set C$ as an ultimate
upper bound to the RA error. Then, subsequently we improve by using the full potential of
$\set V$.  We
first exploit that whenever $\max_{\pi}\psi\geq1-\epsilon$, for some $\epsilon\leq\frac{1}{2}$, then
by Lemma \ref{lemma:RAvsID} the error can be uniformly bounded from above  by
$\frac{\tilde{\lambda}\epsilon}{1-\tilde{\lambda}\epsilon}
=\frac{\lambda^{2}\epsilon}{1+\lambda^{2}\left(  1-\epsilon\right)}
\leq\frac{\lambda^{2}\epsilon}{1+\lambda^{2}\epsilon}$,
and since clearly $\max_{\pi}\phi\geq\frac{1}{n_{t}}$
and $1-\epsilon\leq\frac{\epsilon}{n_{t}-1}$ for
$\epsilon\leq1-\frac{1}{n_{t}}$ we have for $\max_{\pi}\psi\geq\max\left(  0,1-\epsilon\right)$
$
\frac{\lambda^{2}\epsilon}{1+\lambda^{2}\epsilon} 
\leq\frac{\lambda^{2}\epsilon}{1+\lambda^{2}\frac{\epsilon}{n_{t}-1}}
=\frac{\tilde{\lambda}\epsilon}{1+\tilde{\lambda}\left(
\frac{\epsilon}{n_{t}-1}-1\right)  }
$
, for any $\epsilon>0$ (even that for $\epsilon>1$). On the other hand, we have
 $\max_{\pi}\psi<\max\left(  0,1-\epsilon\right)$
$
\frac{|\tilde{\lambda}\psi-\tilde{\vartheta}\phi| }{1-\tilde{\lambda}\psi}
\leq\frac{|\tilde{\lambda}\psi-\tilde{\vartheta}\phi|}{1-\tilde{\lambda}+\tilde{\lambda}\epsilon}
\leq\frac{|\tilde{\lambda}\psi-\tilde{\vartheta}\phi|}{1+\tilde{\lambda}(\frac{\epsilon}{n_{t}-1}-1)}
$
. Hence, we can write for some pair $(\vec{h},\vec{\nu})$:
$
\frac{|\tilde{\lambda}\psi-\tilde{\vartheta}\phi|}{1-\tilde{\lambda}\psi}
\leq\frac{\max\{\tilde{\lambda}\epsilon,|\tilde{\lambda}\psi-\tilde{\vartheta}\phi|\}}{1+
\tilde{\lambda}(\frac{\epsilon}{n_{t}-1}-1)  }
$
and setting
$
\tilde{\lambda}\epsilon   =\min_{\multi{1>\tilde{\vartheta}>0}{\vec{\nu}\in{\mathcal{V}}}}{
\max_{\pi}}|\tilde{\lambda}\psi-\tilde{\vartheta}\phi|
=|\tilde{\lambda}\psi^\ast-
\tilde{\vartheta}^{\ast}\phi^\ast|
$
, where $\phi^\ast=|\langle\vec{\nu}^\ast,\vec{w}_{\pi^\ast(m)}\rangle|^2$ and
$\psi^\ast=|\langle\vec{h},\vec{w}_{\pi^\ast(m)}\rangle|^2$ with respect to maximizing mapping $\pi^\ast$
and minimizing arguments $(\vec{\nu}^\ast,\tilde{\vartheta}^\ast)$. This yields
$
\min_{\multi{1>\tilde{\vartheta}>0}{\vec{\nu}\in{\mathcal{V}}}}{
\max_{\pi}}\frac{|\tilde{\lambda}\psi-\tilde{\vartheta}\phi|}{1-\tilde{\lambda}\psi}
\leq\frac{|\tilde{\lambda}\psi^\ast-\tilde{\vartheta}^{\ast}\phi^\ast|}{1-\tilde{\lambda}+
\frac{1}{n_{t}-1}|\tilde{\lambda}\psi^\ast-\tilde{\vartheta}^{\ast}\phi^\ast|}.
$
Equivalently, for the second term on the right side of (\ref{eq:ExpRVQ}) we have
$
\frac{|\tilde{\lambda}\psi-\tilde{\vartheta}\phi|}{1-\tilde{\vartheta}\phi}
  \leq\frac{\max\{\tilde{\lambda}\epsilon,|\tilde{\lambda}\psi-\tilde{\vartheta}\phi|\}
}{1-\tilde{\lambda}\psi+\tilde{\lambda}\psi-\tilde{\vartheta}\phi}
\leq\frac{\max\{\tilde{\lambda}\epsilon,|\tilde{\lambda}\psi-\tilde{\vartheta}\phi|\}}{1+
\tilde{\lambda}_{m}\left(
\frac{\max\{\epsilon-|\tilde{\lambda}\psi-\tilde{\vartheta}\phi|,0\}  }{n_{t}-1}-1\right)  }
$
. Setting $\epsilon=\left(1+\epsilon^{\prime}\right)
|\tilde{\lambda}\psi^\ast-\tilde{\vartheta}^\ast\phi^\ast|$, $\epsilon^{\prime}>0,$  since the error
term is still increasing in $\epsilon$,  yields
$
\min_{\multi{1>\tilde{\vartheta}>0}{\vec{\nu}\in{\set{V}}}}\max_{\pi}
\frac{|\tilde{\lambda}\psi-\tilde{\vartheta}\phi|}{1-\tilde{\vartheta}\phi}
\leq\frac{\left(  1+\epsilon^{\prime}\right)
|\tilde{\lambda}\psi^\ast-\tilde{\vartheta}^\ast\phi^\ast|}{1-
\tilde{\lambda}+\frac{\epsilon^{\prime}}{n_{t}-1}
|\tilde{\lambda}\psi^\ast-\tilde{\vartheta}^\ast\phi^\ast|},
$
Finally, expanding the fraction with $(1-\tilde \lambda)$ and applying Jensen's inequality again
proves the claim.
\end{IEEEproof}

\section{Proof of Lemma \ref{lem:d(b)}}

\label{proof:d(b)} \begin{IEEEproof}
Consider an arbitrary but fixed user $m\in\set U$ and, without loss of
generality, assume the transmit codebook $\set C$ is given by the standard
ONB.  Define the vector \\ $\vec \psi_m =
\left(|\langle\vec{h}_m,\vec{w}_{\pi(1)}\rangle|^2 , \ldots ,
|\langle\vec{h}_m,\vec{w}_{\pi(n_t)}\rangle|^2\right)$. Since,
$\|\vec{h}_m\| = 1$ from the Parseval's identity follows that $\|\vec
\psi_m\|_1 = 1$ and, therefore, any $\vec \psi_m$ corresponds to a point on
the $d:= n_t-1$ dimensional simplex defined by $\set K_{d} = \{\vec x\in
\Reals^{d+1}: x_i>0,\, i=1,\ldots,n_t \text{ and } \|\vec x \|_1 = 1 \} $,
which has edge length $\sqrt{2}$.  Similarly, each element of the feedback
codebook $\vec \nu \in \set V$ defines a point $\vec q\in \set K_{d}$ on the
$d$--simplex, collected in the set $\set Q \subset \set K_{d}$, with $|\set
Q| =|\set V|=2^B$.  Setting $\tilde \vartheta_m = \tilde \lambda_m $ the
function $D_m(B)$ can be bounded from above by
\[
D_m(B) \leq \Exx{H}{\lambda_m^2}\max_{\vec x \in \set K_{d}} \min_{\vec q
\in \set Q} \|\vec x - \vec q \|_{\infty}.
\]
Now we show that $\delta:=\max_{\vec x \in \set K_{d}} \min_{\vec q \in \set
Q} \|\vec x - \vec q \|_{\infty}$ can be bounded from above by
$c(n_t)2^{-\frac{B}{(n_t-1)}}$, with $c(n_t)$ a constant.
Consider the cubes $\set B_\infty^{n_t}(\vec y ;
\delta):=\{\vec{x}\in\Reals^{n_t} : \lVert\vec y - \vec{x}\rVert_\infty\leq
\delta\}$. If $\vec y \in \set K_d$, then the intersection of the centered
cubes $\set B_\infty^{n_t}(\vec y ; \delta)$ with $\set K_d$ is a polytope
with $2d$ facets. Let $\set
B_2^d(\delta):=\{\vec{x}\in\Reals^d:\lVert\vec{x}\rVert_2\leq \delta \}$ be
the balls with radius $\delta$ that are inscribed in this polytopes.  Hence,
to upperbound the number of centered cubes $\set B_\infty^{n_t}(\vec y ;
\delta)$, with $\vec y \in \set K_d$, required to cover the simplex $\set
K_d$ we need to compute the number of balls $\set B_2^d(\delta)$ required to
cover $\set K_d$.
Let the number of balls $ \set B_2^d(\delta)$ required to cover the simplex
$\set K^d$ be given by the covering number $N(\set K_d, \set
B^d_2(\delta))$. The covering number $N(\set A, \set B)$ is defined as the
number of convex bodies $\set B$ in $\Reals^d$ required to cover a convex
body $\set A$ in $\Reals^d$.  Using the Rogers--Zong Lemma \cite{Rogers1997}
the covering number can be bounded from above by
\begin{equation}\label{eq:Rogers-Zong}
N(\set A,\set B) \leq \Theta(\set B) \frac{\text{vol}(\set A - \set B)}{\text{vol}(\set B)},
\end{equation}
where $\text{vol}(\cdot)$ is a function that computes the volume and
$\Theta(\set B) \geq 1$ is the covering density of $\set B$; if $\Reals^d$
can be tiled by translates of $\set B$ then $\Theta(\set B)=1$; if the
covering has some overlap then $\Theta(\set B)>1$.  Now we can use the
Rogers-Shephard inequality \cite{Rogers1957}, which states that
\begin{equation}\label{eq:Rogers-Shephard}
\text{vol}(\set A - \set B)\text{vol}(\set A  \cap \set B)\leq \binom{2d}{d} \text{vol}(\set A)\text{vol}(\set B).
\end{equation}
Assuming that $\text{vol}(\set A \cap \set B) = \text{vol}(\set B)$ we get
from \eqref{eq:Rogers-Zong} and \eqref{eq:Rogers-Shephard} that the covering
number $N(\set A, \set B)$ is upper bounded by
\begin{equation}
\label{eq:A=B}
N(\set A, \set B) \leq \Theta(\set B) \binom{2d}{d}\frac{\text{vol}(\set
A)}{\text{vol}(\set B)}.
\end{equation}
Now we can apply this bound to our problem.  The volumes of the $d$--simplex
$\set K_d$ and the balls $ \set B^d_2(\delta)$ are $ \text{vol}(\set K_d) =
\frac{\sqrt{d+1}}{d!}$ and $\text{vol}(\set B^d_2(\delta)) =
\frac{\pi^{d/2}}{\Gamma(1+d/2)}\delta^d$, where $\Gamma(\cdot)$ is the gamma
function.  Hence, the covering number can be bounded from above by
\begin{equation*}
N(\set K_d, \set B^d_2( \delta )) = N(\frac{1}{\delta}\set K_d, \set
B^d_2(1))  \leq \Theta(\set B^d_2(1)) \binom{2d}{d} \frac{\Gamma(1+d/2)
\sqrt{d+1}}{d! \pi^{d/2}}\cdot \frac{1}{\delta^d}.
\end{equation*}
Solving for $\delta$ and using $2^B \leq N(\set K_d, \set B^d_2( \delta ))=$
we get
\[
\delta \leq \left(\Theta(\set B^d_2) \binom{2n_t - 2}{n_t-1}
\frac{\Gamma(1+\frac{n_t-1}{2}) \sqrt{n_t}}{(n_t-1)!
\pi^{\frac{n_t-1}{2}}}\right)^{\frac{1}{n_t-1}} 2^{-\frac{B}{n_t-1}}.
\]
Finally, for \eqref{eq:A=B} to be valid we need to ensure that $\text{vol}(\set K_d \cap \set
B_2^d(\delta)) = \text{vol}(\set B_2^d(\delta))$ or, in other words,
$\delta$ is smaller than the inradius of the inscribed circle of the
simplex. According to Klamkin \cite{Klamkin1979} for a regular simplex the
inradius equals the circumradius divided by $n_t-1$. The circumradius is
easily shown by the volume ratio and Stirlings formula to be greater than
$\sqrt{n_t-1}$. 
\end{IEEEproof}

\section{Proof of Theorem \ref{theo:US}}

\label{proof:US} \begin{IEEEproof} Similarly to Lemma \ref{lemma:RAvsID} the
  terms of the sum in \eqref{eq:deltaRA} can be bounded from above.  We use
  again $\phi_{ml}:=|\langle\vec{\nu}_m,\vec{w}_{\pi(l)}\rangle|^2$ and
  $\psi_{ml}:=|\langle\vec{h}_m,\vec{w}_{\pi(l)}\rangle|^2$ without writing
  explicitly the dependency on the mapping $\pi$ and get
  \begin{align*}
    r_{m}(\pi,\set S,\lambda_{m}\vec{h}_{m})-r_{m}(\pi,\set
      S,\vartheta_{m}
      \vec{\nu}_{m}) &=\log\left( \frac{\frac{|\mathcal{S}|}{n_{t}}+\lambda_{m}^{2}{\sum
          _{l\in\mathcal{S}}\psi_{ml}}}{\frac{|\mathcal{S}|}{n_{t}}+
        \lambda_{m}^{2}{\sum_{l\in\mathcal{S}\setminus\{m\}}\psi_{ml}}}\right)
    -\log\left(
      \frac{\frac{|\mathcal{S}|}{n_{t}}+\vartheta_{m}^{2}{\sum
          _{l\in\mathcal{S}}\phi_{ml}}}{\frac
        {|\mathcal{S}|}{n_{t}}+\vartheta_{m}^{2}{\sum_{l\in\mathcal{S}\setminus
            \{m\}}\phi_{ml}}}\right) \\
    & =\log\left( \frac{\frac{|\mathcal{S}|}{n_{t}}+\lambda_{m}^{2}{\sum
          _{l\in\mathcal{S}}\psi_{ml}}}{\frac{|\mathcal{S}|}{n_{t}%
        }+\vartheta_{m}^{2}{\sum_{l\in\mathcal{S}}\phi_{ml}}%
      }\right) +\log\left(
      \frac{\frac{|\mathcal{S}|}{n_{t}}+\vartheta_{m}%
        ^{2}{\sum_{l\in\mathcal{S}\setminus\{m\}}\phi_{ml}}}%
      {\frac{|\mathcal{S}|}{n_{t}}+\lambda_{m}^{2}{\sum_{l\in\mathcal{S}%
            \setminus\{m\}}\psi_{ml}}}\right).
  \end{align*}
  Setting $\vartheta_{m}^{2}=\lambda_{m}^{2}$ we get the inequality chain
  \begin{align*}
   r_{m}(\pi,\set S,\lambda_{m}\vec{h}_{m})-r_{m}(\pi,\set
      S,\lambda_{m}
      \vec{\nu}_{m}) &=\log\left( 1+\frac{n_{t}{\lambda_{m}^{2}}}{|\mathcal{S}|}\frac
      {{\sum_{l\in\mathcal{S}}\psi_{ml}}-\phi_{ml}}{1+
        \lambda_{m}^{2}{\sum_{l\in\mathcal{S}}\phi_{ml}} }\right) 
    +\log\left( 1+\frac{n_{t}{\lambda_{m}^{2}}}{|\mathcal{S}|}%
      \frac{{\sum_{l\in\mathcal{S}\setminus\{m\}}\phi_{ml}}%
        -\psi_{ml}}{1+\lambda_{m}^{2}{\sum_{l\in\mathcal{S}%
            \setminus\{m\}}\psi_{ml}}}\right) \\
    & \leq\log\left( 1+\frac{n_{t}{\lambda_{m}^{2}}}{|\mathcal{S}|}{\left\vert
          \sum_{l\in\mathcal{S}}\psi_{ml}-\phi_{ml}\right\vert }\right)+\log\left( 1+\frac{n_{t}{\lambda_{m}^{2}}%
      }{|\mathcal{S}|}{\left\vert
          \sum_{l\in\mathcal{S}\setminus\{m\}}\psi_{ml}-
          \phi_{ml}\right\vert }\right) \\
    & \leq\log\left( 1+\frac{n_{t}{\lambda_{m}^{2}}}{|\mathcal{S}|}\left\vert
        \mathcal{S}\right\vert \max_{\pi}\left\vert \psi_{mm}-
        \phi_{mm}\right\vert \right)  +\log\left( 1+\frac
      {n_{t}{\lambda_{m}^{2}}}{|\mathcal{S}|}\left\vert \mathcal{S}\setminus
        \{m\}\right\vert \cdot\max_{\pi}\left\vert \psi_{mm}-\phi_{mm}\right\vert \right) \\
    & \leq2\log\left( 1+n_{t}{\lambda_{m}^{2}}\max_{\pi}\left\vert
        \psi_{mm}-\phi_{mm}\right\vert \right) \\ & =2\log\left(
      1+\frac{P\|\vec{\hat h}_m\|_2^{2}}{\sigma^{2}}\max_{\pi}\left\vert
        \psi_{mm}-\phi_{mm}\right\vert \right) .\nonumber
  \end{align*}
  The lower bound on $-(r_{m}(\pi,\set S,\lambda_{m}\vec{h}_{m}%
  )-r_{m}(\pi,\set S,\vartheta_{m}\vec{\nu}_{m}))$ can be obtained in a
  similar manner. Taking expectations and using Jensen's inequality we obtain
  \begin{equation*}
    \Exx{H} {r_{m}(\pi,\set S,\lambda_{m}\vec{h}_{m})-r_{m}(\pi,\set
      S,\lambda_{m} \vec{\nu}_{m})}  \leq 2\log\left( 1+\Exx{\vec
        H}{\frac{P\|\vec{\hat h}_m\|_2^{2}}{\sigma^{2}}\max_{\pi}\left\vert
          \psi_{mm}-\phi_{mm}\right\vert }\right).
  \end{equation*}
  Since $\max_{\pi}|\psi_{mm}-\phi_{mm}|$ depends only on the channel
  directions $\vec h_m$, it is independent of the channel magnitude
  $\|\vec{\hat h}_m\|_2$.
  \begin{equation*}
    \Exx{H} {r_{m}(\pi,\set S,\lambda_{m}\vec{h}_{m})-r_{m}(\pi,\set
      S,\lambda_{m} \vec{\nu}_{m})}  \leq 2\log\left( 1+\frac{P
        n_t}{\sigma^{2}}\Exx{H}{\max_{\pi}\left\vert
          \psi_{mm}-\phi_{mm}\right\vert }\right).
  \end{equation*}
  Using the RA scheme and \eqref{eq:deltaRA} yields the result.
\end{IEEEproof}

\bibliographystyle{IEEEtran} \bibliography{IEEEabrv,refs}

\begin{IEEEbiography}
  {Gerhard Wunder} (M'05) studied electrical engineering at the University of
  Hannover, Germany, and the Technische Universit\"at (TU) Berlin, Germany, and
  received his graduate degree in electrical engineering (Dipl.-Ing.) with
  highest honors in 1999 and the PhD degree (Dr.-Ing.) in communication
  engineering on the peak-to-average power ratio (PAPR) problem in OFDM with
  distinction (summa cum laude) in 2003 from TU Berlin. In 2007, he also
  received the habilitation degree (venia legendi) and became a Privatdozent
  at the TU Berlin in the field of detection/estimation theory, stochastic
  processes and information theory. Since 2003 he is heading a research group
  at the Fraunhofer Lab for Mobile Communications (FhG-MCI) working in close
  collaboration with industry on theoretical and practical problems in
  wireless communication networks particularly in the field of LTE-A
  systems. He is a recipient of research fellowships from the German national
  research foundation.

  In 2000 and 2005, he was a visiting professor at the Georgia Institute of
  Technology (Prof. Jayant) in Atlanta (USA, GA), and the Stanford University
  (Prof. Paulraj) in Palo Alto/USA (CA). In 2009 he was a consultant at
  Alcatel-Lucent Bell Labs (USA, NJ), both in Murray Hill and Crawford
  Hill. He was a general co-chair of the 2009 International ITG Workshop on
  Smart Antennas (WSA 2009) and a lead guest editor in 2011 for a special
  issue of the Journal of Advances on Signal Processing regarding the PAPR
  problem of the European Association for Signal Processing. Since 2011, he is
  also an editor for the IEEE Transactions on Wireless Communications
  (TWireless) in the area of Wireless Communications Theory and Systems
  (WCTS). In 2011 Dr. Wunder received the best paper award for outstanding
  scientific publication in the field of communication engineering by the
  German communication engineering society (ITG Price 2011).
\end{IEEEbiography}

\begin{IEEEbiography}
  {Jan Schreck} (Member IEEE) received the Dipl.-Math. in 2006 from the University of
  Applied Science Berlin, Germany, in cooperation with the Weierstrass
  Institute for Applied Analysis and Stochastics Berlin, Germany. 
  Since 2006 he has been with the Fraunhofer German-Sino Lab for Mobile
  Communications and since 2011 he is with the Technische Universit\"at
  Berlin, Germany, where he is currently working toward the Ph.D. degree. 
  His research interests include information theory and signal
  processing. 
\end{IEEEbiography}

\begin{IEEEbiography}
  {Peter Jung} (Member IEEE, Member VDE/ITG) received the Dipl.-Phys. in high
  energy physics in 2000 from Humboldt University, Berlin, Germany, in
  cooperation with DESY Hamburg. Since 2001 he has been with the Department of
  Broadband Mobile Communication Networks, Fraunhofer Institute for
  Telecommunications, Heinrich-Hertz-Institut (HHI) and since 2004 with
  Fraunhofer German-Sino Lab for Mobile Communications.  He received the
  Dr.-rer.nat (Ph.D.) degree in 2007 (on Weyl--Heisenberg representations in
  communication theory) at the Technical University of Berlin (TUB),
  Germany. P. Jung is currently working under DFG grants JU 2795\/1-1\&2 at
  the Technical University in Berlin, Germany (TUB) and Technical University
  in Munich, Germany (TUM) in the field information theory and signal
  processing.  His current research interests are in the area time--frequency
  analysis, compressed sensing, dimension reduction and randomized algorithms.
  He is giving lectures in compressed sensing and estimation theory.
\end{IEEEbiography}

\end{document}